\documentclass[%
reprint,
superscriptaddress,
 amsmath,amssymb,
 aps,
prb,
]{revtex4-2}

\usepackage{graphicx}
\usepackage{float}
\usepackage{dcolumn}
\usepackage{bm}
\usepackage{mathtools, amsmath, amsthm, amssymb, amsfonts}
\usepackage{nicefrac}
\usepackage[intlimits]{esint}
\usepackage{bbold}
\usepackage{xcolor}

\begin{document}

\preprint{APS/123-QED}

\title{First principles study of topological invariants of Weyl points in continuous media}

\author{G. R. Fonseca}
\email{guilhermefonseca@tecnico.ulisboa.pt}
\affiliation{
Instituto de Telecomunicações, Instituto Superior Técnico-University of Lisbon, Avenida Rovisco Pais 1, Lisboa, 1049-001 Portugal
}
\author{F. R. Prudêncio}%
\affiliation{
Instituto de Telecomunicações, Instituto Superior Técnico-University of Lisbon, Avenida Rovisco Pais 1, Lisboa, 1049-001 Portugal
}
\affiliation{%
Instituto Universitário de Lisboa (ISCTE-IUL), Avenida das Forças Armadas 376, 1600-077 Lisbon, Portugal
}%

\author{M. G. Silveirinha}%
\affiliation{
Instituto de Telecomunicações, Instituto Superior Técnico-University of Lisbon, Avenida Rovisco Pais 1, Lisboa, 1049-001 Portugal
}


\author{P. A. Huidobro}
\email{p.arroyo-huidobro@uam.es}
\affiliation{
Instituto de Telecomunicações, Instituto Superior Técnico-University of Lisbon, Avenida Rovisco Pais 1, Lisboa, 1049-001 Portugal
}
\affiliation{Departamento de F\'{i}sica Te\'orica de la Materia Condensada and Condensed Matter Physics Center (IFIMAC), Universidad Aut\'onoma de Madrid, E-28049 Madrid, Spain
}%


\date{\today}

\begin{abstract}
In recent years there has been a great interest in topological photonics and protected edge states. Here, we present a first principles method to compute topological invariants of three-dimensional gapless phases. Our approach allows to calculate the topological charges of Weyl points through the efficient numerical computation of gap Chern numbers, which relies solely on the photonic Green's function of the system. We particularize the framework to the Weyl points that are found to emerge in a magnetized plasma due to the breaking of time reversal symmetry. We discuss the relevance of modelling nonlocality when considering the topological properties of continuous media such as the magnetized plasma. We find that for some of the considered material models the charge of the Weyl point can be expressed in terms of a difference of the gap Chern numbers of two-dimensional material subcomponents. Our theory may be extended to other three-dimensional topological phases, or to Floquet systems.
\end{abstract}

\maketitle


\section{Introduction}

Topological states in photonics have attracted much attention due to the promise of light propagation protected from backscattering and thus immune to defects, such as sharp edges or corners of the medium surface \cite{PhysRevA.78.033834,PhysRevLett.100.013904,RevModPhys.91.015006,Paloma,Ling,toporecent}. 
Through the bulk-edge correspondence, a non-trivial topological invariant of the medium's bands guarantees topologically protected edge modes that lie in the band gap, when the topological medium is interfaced with a topologically trivial material, or vacuum \cite{PhysRevB.94.205105,PhysRevX.9.011037}.  
The discovery of nontrivial topological phases in optics may be traced back to the study of photonic crystals \cite{PhysRevA.78.033834,PhysRevLett.100.013904,Soljacic} inspired by the electronic models such as the quantum Hall effect \cite{PhysRevLett.71.3697,PhysRevB.84.125132,PhysRevLett.95.146802}. The non-trivial topological invariants of these models are non-zero, integer Chern numbers. Topological edge modes have also been observed in a number of other systems including waveguide arrays \cite{Rechtsman2013}, but also in photonic continua \cite{Khanikaev2013,PhysRevB.92.125153,Mario2016Z2,PhysRevApplied.12.014021,David}. 

Three-dimensional (3D) systems can also host gapless topological phases. This is the case of Weyl semimetals, characterized by point degeneracies between topologically inequivalent bands. These linear crossings between bands in 3D, named Weyl points, are robust against perturbations and are characterized by a quantized topological charge \cite{toporecent}.  
Weyl points have been realised in complex 3D photonic systems, such as photonic crystals \cite{gyroid,electronWeyl} and metamaterials \cite{PhysRevLett.117.057401,MetaWeyl,PhysRevLett.123.033901} as well as in homogeneous continuous media such as a magnetized plasma \cite{ncomms12435}. The associated protected surface states that emerge in the gaps away from the Weyl point have been observed in photonic systems including a strongly biased semiconductor \cite{InSbWeyl} and a metamaterial \cite{Kim2019Meta}. 


Topological photonic systems are conventionally characterized by means of topological band theory \cite{Ling,toporecent,dePaz2020}. This framework yields a band's Chern number from an integral of the Berry curvature in momentum space, which is in turn calculated from the system's eigenstates. This approach is usually taken for periodic photonic crystals, with momentum space integrals carried out in the Brillouin zone, and typically neglects dispersion, which is on the other hand an essential ingredient in topological continuous media, such as the magnetized plasma.   
An alternative approach to characterize Weyl points in the past has relied on linear $k\cdot p$ models around the Weyl point, which enables one to write a Hamiltonian for the system \cite{ncomms12435}. 

In this work, we introduce a first principles method to calculate topological invariants in 3D dispersive photonic systems, and we apply it to the study of the Weyl points in a magnetized plasma. 
In contrast to topological band theory, our approach is based on the Green's function and does not require the calculation of the system eigenstates, the direct computation of Berry curvature or the bands' Chern numbers \cite{PhysRevB.97.115146}. 
Additionally, dispersion is naturally included in the method, as well as non-locality. In fact, it is known that a non-local material response must be taken into account in order to guarantee well-defined topologies in dispersive photonic systems \cite{PhysRevB.92.125153,PhysRevB.94.205105,Hanson,PhysRevLett.124.153901,PhysRevB.105.035310,PhysRevLett.129.133903}. Here, we find that nonlocality can also influence in a decisive way the number and the type of Weyl points in an electromagnetic continuum.

The paper is organised as follows. First, in section II we detail the methodology that allows us to compute the topological charges of Weyl points from first principles, based on calculating the gap Chern numbers in two-dimensional (2D) cross sectional planes away from the Weyl point under study. Next, in section III we present the results for the topological charge of Weyl points in a magnetized plasma. We first consider a local model which is known to yield non-integer topological invariants in 2D type systems \cite{PhysRevB.92.125153}. In order to take into account the unavoidable nonlocality of the magnetized plasma, we then consider two models that incorporate a physical regularization procedure in the material response: a non-local hydrodynamic model and a full spatial cut-off model. We discuss the Weyl points that emerge in both systems, as well as their differences.
 

\section{Analytical Formalism}
Weyl points are topological entities that originate from a geometric property of a system's eigenstates, the Berry curvature. Integration of the Berry curvature, $\boldsymbol{\mathcal{F}}_n (\textbf{k})$, over a closed surface that encloses the point gives a quantized quantity, 
\begin{equation}
    \int \text{d}S \, \textbf{n}\cdot \boldsymbol{\mathcal{F}}_n (\textbf{k}) = 2\pi \mathcal{C}_W \label{eq:CW_Berry}
\end{equation}
with $\textbf{n}$ a unit vector normal to the integration surface in the outgoing direction and $n$ and index that labels the bands. $\mathcal{C}_W$, an integer, is the charge of the Weyl point. Its sign gives the chirality of the Weyl degeneracy, that is, if the flux is incoming or outgoing. Thus, the amount of Berry flux that goes through any closed surface in momentum space that encloses the Weyl point is a topological invariant. This implies that Weyl points are monopoles of Berry curvature in 3D momentum space. They only exist in systems with a broken parity ($\mathcal{P}$) or time reversal ($\mathcal{T}$) symmetries, or both, which explains their robustness against perturbations. In fact, they always appear in pairs, with one source and one drain, that can mutually annihilate. 

\begin{figure}
    \centering
    \includegraphics[width=0.5\textwidth]{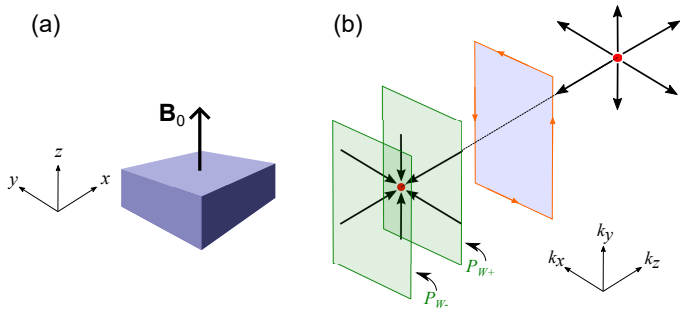}
    \caption{Weyl points are monopoles of Berry curvature in momentum space. A plasma magnetized in the $\mathbf{\hat{z}}$ direction (a) hosts pairs of Weyl points aligned along the $k_z$ axis (b). 
    A slice of momentum space between two Weyl nodes is a Chern insulator (blue plane) and hosts unidirectional edge states. The planes in green on the right $(P_{W+})$ and on the left $(P_{W-})$ of the Weyl point represent the cuts where the gap Chern numbers will be numerically computed and whose difference gives the topological charge of the Weyl point.}
    \label{Weyl}
\end{figure}

Let us now assume a system with broken $\mathcal{T}$, where a minimum of two Weyl points appear at opposite momenta, as sketched in Fig. \ref{Weyl}. This is for the sake of simplicity, and a similar procedure can be applied to systems with broken $\mathcal{P}$ which host a minimum of two pairs of Weyl points. Consider the topological charge of one of the two Weyl points of a pair, $\mathcal{C}_W$ in Eq. \ref{eq:CW_Berry}. The surface integral in that equation can be carried out in an infinitesimally small sphere enclosing the Weyl point. By construction, inside the integration volume the bands are well separated by a band gap and the Berry curvature is well defined for the bands of interest everywhere except at the Weyl point. Thus the divergence of the Berry curvature is zero everywhere except at the singularity, and through Stokes theorem the integration surface can be deformed into a parallelepiped. Thus the integral can be equivalently carried out on the surface of a rectangular slice of momentum space perpendicular to the axis along which the Weyl point pairs are aligned (the $k_z$ axis in the sketch of Fig. \ref{Weyl}b). Then, the topological invariant amounts to a sum of the Berry fluxes going through all the sides of this parallelepiped. Furthermore, the choice of integration surface ensures that there are no other sources of Berry flux enclosed by it, such that we can take the side faces (parallel to the $k_{z}$ axis) to infinity where their contribution vanishes, and obtain the topological invariant from the Berry fluxes going through the two planes orthogonal to the $k_z$ axis at either side of the Weyl point, $P_{W-}$ and $P_{W+}$, sketched in green in Fig. \ref{Weyl}b. Thus, the expression for the topological charge is,
\begin{equation} \label{eq:CW_planes}
    \mathcal{C}_W = \frac{1}{2\pi} \int_{P_{W-}} \text{d}^2\textbf{k}  \, (-\mathbf{\hat{z}})\cdot \boldsymbol{\mathcal{F}}_{n} (\textbf{k}) + \frac{1}{2\pi} \int_{P_{W+}} \text{d}^2\textbf{k} \, \mathbf{\hat{z}}\cdot\boldsymbol{\mathcal{F}}_{n} (\textbf{k}),
\end{equation}
with $\mathbf{\hat{z}}$ serving as the unit vector along $k_{z}$, and where the relative sign between the two terms takes into account that the unit vecor $\textbf{n}$ in Eq. (1) points outwards to the parallelepiped's surface.
Additionally, the label $n$ corresponds to the band that lies below the Weyl point. 
Conventionally, the Berry curvature is calculated from the Berry connection, $\boldsymbol{\mathcal{A}}_n (\textbf{k})$, as $\boldsymbol{\mathcal{F}}_n (\textbf{k}) = \nabla_\textbf{k} \times\boldsymbol{\mathcal{A}}_n (\textbf{k})$, with $\boldsymbol{\mathcal{A}}_n (\textbf{k}) = i \int \text{d}^3\textbf{k} \, \Psi^\dagger_n (\textbf{k}) \nabla_\textbf{k}  \Psi_n (\textbf{k}) $, which requires knowledge of the eigenfunctions along the 3D bands $\Psi_n (\textbf{k})$. 

Here we take an alternative approach that circumvents the computation of the eigenfunctions, by observing that the integrals in Eq. \ref{eq:CW_planes} correspond to the Chern numbers of a given band at each side of the Weyl point, in particular, the bands that lie below it. Furthermore, since the bands only touch at the Weyl point and there are well defined band gaps at each side of it, the Chern numbers of the bands can be replaced by \textit{gap} Chern numbers,  
\begin{eqnarray}
    \mathcal{C}_{W} 
    =\mathcal{C}_\text{gap}(P_{W+})-\mathcal{C}_\text{gap}(P_{W-})
    \label{charge}
\end{eqnarray} 
with $\mathcal{C}_\text{gap}=\sum_{n}\mathcal{C}_{n}$ defined as the sum of the individual Chern numbers of the bands immediately below the given band gap. This quantity can be calculated from first principles using the photonic Green's function of the medium, as was shown in Refs. \cite{PhysRevB.97.115146,Chern,sym13112229,PhysRevLett.129.133903}.

The photonic Green's function is given by:
\begin{equation}
    \mathcal{G}_{\mathbf{k}}=i\left(\hat{L}_{\mathbf{k}}-\omega\mathbf{1}\right)^{-1},
    \label{Green}
\end{equation}
where $\mathbf{1}$ is the identity and $\hat{L}_{\mathbf{k}}$ is a frequency-independent differential operator that effectively models the propagation in a dispersive medium. Typically, this entails modeling the effects of the material dispersion with additional variables that represent the internal degrees of freedom of the medium responsible for the dispersive response. The operator $\hat{L}_{\mathbf{k}}$ is parameterized by the real wave vector $\mathbf{k}$ and and its eigenfrequencies coincide with the poles of $ \mathcal{G}_{\mathbf{k}}$. Its derivation is presented in Appendix A. For lossless media, the eigenfrequencies lie in the real frequency axis, separated by vertical strips that correspond to the band gaps. The gap Chern number of each spectral band gap can be expressed in terms of the Green’s function through an integral in the complex frequency space over a line parallel to the imaginary axis contained in the band gap ($\omega =  \omega_\text{gap}+i\infty$),
\begin{widetext}
\begin{equation}
    \mathcal{C}_\text{gap}=\frac{i}{(2\pi)^{2}}\iint\text{d}^{2}\textbf{k}\, \int^{\omega_\text{gap}+i\infty}_{\omega_\text{gap}-i\infty}\text{d}\omega \, \text{Tr}\{\partial_{1}\hat{L}_{\mathbf{k}}\cdot \mathcal{G}_{\mathbf{k}}\cdot\partial_{2}\hat{L}_{\mathbf{k}}\cdot\mathcal{G}^{2}_{\mathbf{k}}\}.
    \label{Greenformalism}
\end{equation}
\end{widetext}
Here, $\text{Tr}$ is the trace operator, $\partial_i = \partial/\partial k_i$ ($i=x,y$), $\omega_\text{gap}$ is some frequency contained within the band gap, and the integral over $\mathbf{k}$ is carried out over the whole momentum space.

\begin{figure*}[ht!]
    \centering
    \includegraphics[width=0.9\textwidth]{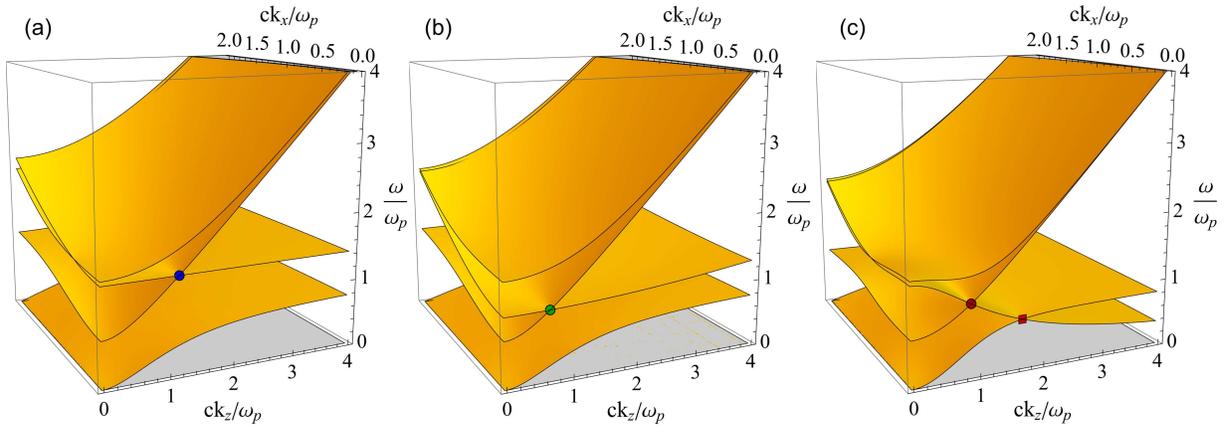}
    \caption{Three-dimensional dispersion of local and nonlocal magnetized plasma models. (a) Dispersion for the local Drude model described in section III.A., with parameters $\omega_{c}=0.8\omega_{p}$ and $\omega'_{p}=\sqrt{2}\omega_{p}$. (b) Dispersion surfaces for a magnetized plasma described by a hydrodynamic model with $\beta=0.2c$ and $\omega_{c}=0.8\omega_{p}$. (c) Frequency dispersion for a magnetized plasma described with a full spatial cut-off model with $k_{max}=\omega_{p}/c$, $\omega_{c}=0.8\omega_{p}$ and $\omega'_{p}=\sqrt{2}\omega_{p}$. }
    \label{disp}
\end{figure*}



Hence, with this method the topological charge of a Weyl point is obtained by computing the gap Chern number of the frequency band gaps that exist at each side of the Weyl point under study. This approach is valid for dispersive continuous media, as well as for photonic crystals \cite{PhysRevLett.129.133903}.
In the following section we apply this framework to study a continuous medium with non-trivial topology featuring Weyl points: a magnetized plasma \cite{ncomms12435}. Three different models are considered, a local and two non-local models, which show that including non-locality is important in the study of topological properties of homogeneous photonic Weyl media.

\section{\label{topo}Topological Characterization of Magnetized Plasma}

Let us consider a 3D plasma biased with a magnetic field applied along the $\mathbf{\hat{z}}$ direction, as sketched in  Fig. \ref{Weyl}(a). The medium is characterized by a plasma frequency $\omega_{p}$ and a cyclotron frequency $\omega_{c}$. Throughout this work, we consider lossless local and nonlocal models of magnetized plasma, with a nonmagnetic response, $\overline{\mu}=\mu_{0}\mathbf{1}$. The nonreciprocal edge modes supported by a magnetized plasma in the plane transverse to the magnetic field are already well studied, taking the 3D medium as translationally invariant and reducing it to a 2D problem \cite{PhysRevB.92.125153,PhysRevB.94.205105,Hanson,PhysRevLett.124.153901,PhysRevB.105.035310,PhysRevLett.129.133903,Chettiar:14}.

Breaking of $\mathcal{T}$-symmetry enables the emergence of Weyl point pairs in this 3D system \cite{ncomms12435}. The Weyl points arise as linear crossings between longitudinal and transverse modes along the $k_{z}$ axis, as imposed by the direction of the applied magnetic field. As $\mathcal{P}$-symmetry is preserved, $\omega(-\mathbf{k})=\omega(\mathbf{k})$, and each Weyl crossing with positive momentum $k_{z}$ has a partner at a symmetric point $-k_{z}$. The 3D dispersion surfaces of magnetized plasmas with different nonlocal properties can be seen in figure \ref{disp}: a local model is considered in (a), and two different non-local models in (b,c), with parameters given in the caption. Weyl crossings at positive $k_z$ can be seen in all cases, marked with a blue, green or red dot in each panel. As mentioned, $\mathcal{P}$-symmetry ensures that band structures are symmetric between $\textbf{k}$ and $-\textbf{k}$, so we only show positive $k_z$. 
As can be seen already in this figure and we discuss in detail below, the topological properties of the three models are not identical, that is, the number of Weyl point pairs, the type of Weyl points, and their topological charge differs depending on the chosen model. This highlights the fact that it is important to take non-locality into account when studying topological properties of continuous media \cite{PhysRevB.92.125153,PhysRevB.94.205105,Hanson,PhysRevLett.124.153901,PhysRevB.105.035310,PhysRevLett.129.133903}.


\subsection{Local Model}

We first discuss the local magnetized plasma. The permittivity tensor that describes electromagnetic propagation in the bulk of this continuum is a matrix with a gyrotropic structure \cite{Kong,PhysRevB.13.2497}:
\begin{equation}
    \frac{\overline{\epsilon}\left(\omega\right)}{\epsilon_{0}}=\mathbf{1}+\frac{\omega_{p}^{2}}{\omega}\left[-\omega\mathbf{1}+i\omega_{c}\mathbf{\hat{z}}\times\mathbf{1}_{t}\right]^{-1}=\begin{bmatrix}\epsilon_{t} & -i\epsilon_{g} & 0 \\ i\epsilon_{g} & \epsilon_{t} & 0 \\ 0 & 0 & \epsilon_{z}\end{bmatrix},
    \label{permittivitytensor}
\end{equation}
where $\mathbf{1}_{t}=\mathbf{\hat{x}}\otimes\mathbf{\hat{x}}+\mathbf{\hat{y}}\otimes\mathbf{\hat{y}}$, and with $\times$ as the cross product and $\otimes$ as the tensor product. Each matrix component is
\begin{equation}
    \epsilon_{t}=1-\frac{\omega^{2}_{p}}{\omega^{2}-\omega^{2}_{c}}, \quad
    \epsilon_{g}=\frac{-\omega_{c}\omega^{2}_{p}}{\omega(\omega^{2}-\omega^{2}_{c})}, \quad
    \epsilon_{z}=1-\frac{\omega^{2}_{p}}{\omega^{2}}.
    \label{tensorentries}
\end{equation}
The lack of transpose symmetry is a consequence of breaking time-reversal symmetry ($\mathcal{T}$) through the application of the magnetic field, turning the plasma into a nonreciprocal medium. 

\begin{figure}[H]
    \centering
    \includegraphics[width=0.50\textwidth]{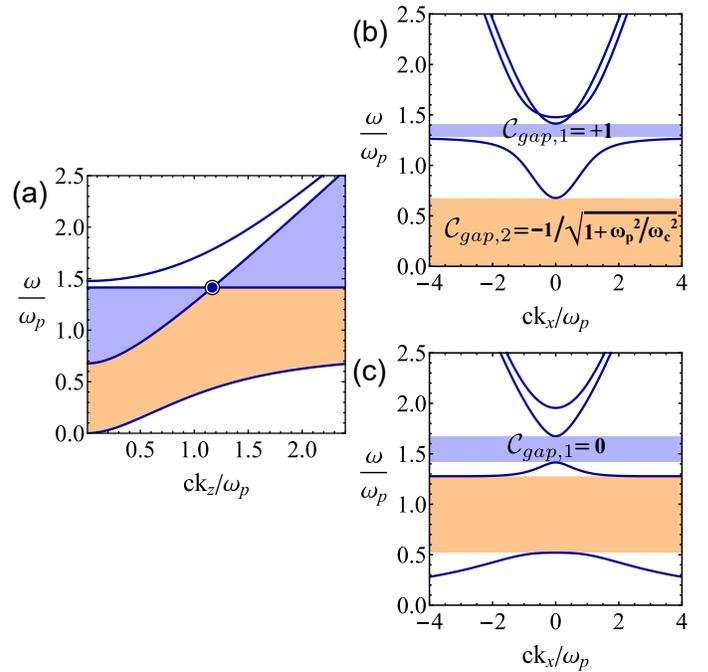}
    \caption{Frequency dispersion for a local magnetized plasma, in cross sections of Fig. \ref{disp}(a). (a) Dispersion along the $k_{z}$ axis exhibiting one of the Weyl points of the pair marked with a blue circle. The band gaps of interest are shaded in blue and orange. (b)-(c) Dispersion in cross-sectional planes orthogonal to the $k_z$ axis, before and after the Weyl point: $k_z=0$ (b) and $k_z=\frac{5}{4}k_{z}^W$ (c). The parameters used here were $\omega_{c}=0.8\omega_{p}$ and $\omega'_{p}=\sqrt{2}\omega_{p}$.}
    \label{local}
\end{figure}

The local model considered in this work has a permittivity tensor $\overline{\epsilon}_{loc}$ which is the same as in equation (\ref{permittivitytensor}), but with a uniaxial plasma response, characterized with a different plasma frequency in the $\mathbf{\hat{z}}$ direction, $\omega'_{p}$.
Only the $\mathbf{\hat{z}}\otimes\mathbf{\hat{z}}$ component is affected by this choice: $\epsilon_{z}=\epsilon_{0}\left(1-\omega'^{2}_{p}/\omega^{2}\right)$. Under these conditions, this model only possesses one Weyl pair and full band gaps are ensured in the $xoy$ plane away from the Weyl points. The Weyl points location is given by:
\begin{equation}
    k_{z} = \pm k_{z}^W=\pm\sqrt{\frac{\omega^{'}_{p}(\omega^{'2}_{p}+\omega_{c}\omega^{'}_{p}-\omega^{2}_{p})}{c^{2}(\omega_{c}+\omega^{'}_{p})}}.
    \label{Weylmomenta}
\end{equation} 
The Weyl points arise along the direction of the magnetic field in momentum space, $k_z$, as a linear crossing between a flat longitudinal plasmon mode with constant frequency $\omega=\omega'_{p}$ and a transverse mode. The Weyl point with positive $k_z$  is highlighted with a blue dot in figure \ref{local}(a), and as previously mentioned the bands are symmetric under $k_z\rightarrow-k_z$ (not shown here). 
As they stem from the flat longitudinal mode, the point-like degeneracies displayed by this model are tilted Weyl points that exhibit parabolic isofrequency surfaces around them \cite{ncomms12435}.
They stand at the critical transition between type-I Weyl points, with ellipsoidal isofrequency surfaces, and type-II Weyl points, for which the isofrequency surface is a hyperboloid \cite{PhysRevLett.117.057401}. 

In order to characterize the topological properties of the medium described by this local model with our methodology, we consider the dispersion in cross sectional planes of the 3D bands orthogonal to the $k_{z}$ axis. As described in section II, we consider planes at the left and right of the Weyl point
and numerically compute the gap Chern numbers of the band gaps that exist in these cuts. 
Figure \ref{local}(b), depicts a cross section of the 3D dispersion at $k_{z}=0$ ($xoy$ plane), which displays two band gaps: a low-frequency one, bounded between zero-frequency mode and a low-frequency transverse magnetic (TM) band (shaded in orange); and a high-frequency gap between the same TM mode and a transverse electric (TE) mode (shaded in blue). 
The Chern numbers of these band gaps have been previously studied \cite{PhysRevB.95.115103,PhysRevB.92.125153,PhysRevB.97.115146}, and analytical expressions have been found (for $k_{z}=0$). The gap Chern numbers take into account the topological charge of negative frequency bands (not shown in Fig. \ref{local}), as they are given by the sum of the Chern numbers of all bands below it. For the high-frequency band gap (highlighted in blue), $\mathcal{C}_{gap,1}=+1$, while for the low frequency one $\mathcal{C}_{gap,2}=-1/\sqrt{1+\omega_{p}^{2}/\omega_{c}^{2}}$. 
The latter result, a non-integer gap Chern number, is a consequence of the ill-defined topology of the low-frequency TM mode due to the limitations of the local model for material parameters. 
Next we take a cross section of the dispersion at the right-hand side of the Weyl point, in particular at $k_z=\frac{5}{4}k_z^W$ (see Fig. 3c). As $k_z$ increases from $k_z=0$, the low frequency TM mode bends upwards and crosses the flat TE mode at the Weyl point. This band crossing enables an exchange of topological charge, and thus modifies the gap Chern number of the high frequency (blue) band gap. We confirm this numerically (Eq. \ref{Greenformalism}) by computing the gap Chern number, which yields $\mathcal{C}_{gap,1}=0$. On the other hand, the low frequency (orange) band gap does not close and thus its topological invariant stays a non-integer number, decreasing as $k_z$ increases. 
Additionally, the results are symmetric for negative momentum values ($k_{z}=-k_{z}^W$), due to the spatial parity symmetry of the Berry curvature $\mathcal{F}(-\mathbf{k})=\mathcal{F}(\mathbf{k})$ \cite{ncomms12435,PhysRevLett.93.206602}. Thus, we have that for $k_z \in (-\infty, -k_z^W)$, $\mathcal{C}_{gap,1}=0$, while for $k_z \in (-k_z^W,k_z^W)$, $\mathcal{C}_{gap,1}=+1$ and for $k_z \in (k_z^W,\infty)$, $\mathcal{C}_{gap,1}=0$. From this we compute the topological charge of the two Weyl points following Eq. (\ref{eq:CW_planes}) as $\mathcal{C}_{W} (k_z=-k_z^W)=(+1)-0=+1$ for the one at negative momentum, and $\mathcal{C}_{W} (k_z=+k_z^W)=0-(+1)=-1$ for the one at positive momentum. This is consistent with the results obtained in Ref. \cite{ncomms12435} for a local magnetized plasma using a $\textbf{k}\cdot\textbf{p}$ expansion close to the Weyl points. As expected, having $\omega'_{p} \neq \omega_{p}$ in the permittivity tensor does not change the topological properties of the magnetized plasma, which shows one pair of Weyl points with charge $\pm1$ for $\omega{c}<\omega_{p}$.

Due to the continuous and local nature of this medium, some of its topological features are ill-defined. To overcome this problem, some cut-off should be included to the material response, so as to ensure that its nonreciprocal components are suppressed for large wave vectors \cite{PhysRevB.92.125153}. When dealing with realistic materials, fields with very fast spatial variation cannot effectively polarize the microscopic constituents of the medium, therefore its response is effectively suppressed when $k\rightarrow\infty$ and it should reduce to that of the vacuum, so the momentum cut-off has a physical justification \cite{landau2013electrodynamics}.
For this purpose, we next consider two different nonlocal models: the hydrodynamic model and the full cut-off model.

\begin{figure}[ht]
    \centering
    \includegraphics[width=0.55\linewidth]{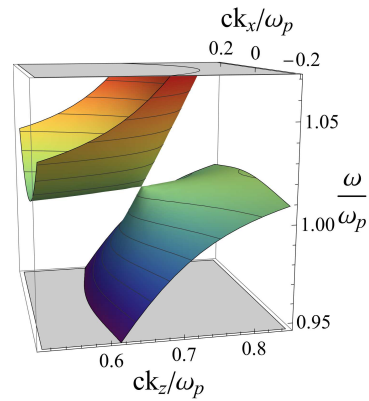}
    \caption{The hydrodynamic model features Type-II Weyl points: Dispersion surfaces and isofrequency curves around the Weyl point at $k_z=+k_{z}^W$ in the $\hat{\mathbf{z}}$ and $\hat{\mathbf{x}}$ directions. The parameters used here were $\beta=0.2c$ and $\omega_{c}=0.8\omega_{p}$.}
    \label{hydroWeyl}
\end{figure}

\begin{figure*}
    \centering
    \includegraphics[width=\textwidth]{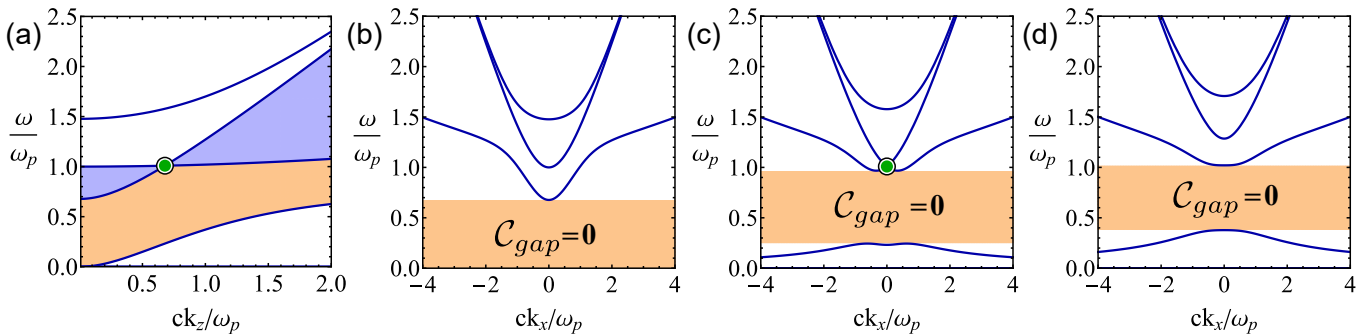}
    \caption{Hydrodynamic model: (a) Dispersion in the $k_{z}$ axis exhibiting the two regions of interest shaded in blue and orange. (b)-(d) Cross sections of the 3D dispersion (FIG. \ref{disp}(b)) for different values of $k_{z}$: (b) $k_{z}=0$, (c) $k_{z}=k_{z}^W$ and (d) $k_{z}=\frac{3}{2}k_{z}^W$. The band gaps are highlighted in orange. The Weyl crossing is highlighted with the green circle. The parameters used here were $\beta=0.2c$ and $\omega_{c}=0.8\omega_{p}$.}
    \label{hydro}
\end{figure*}

\subsection{Hydrodynamic Model}

A conventional approach to consider nonlocal effects is to employ the hydrodynamic model, where the repulsive interactions between electrons are accounted for by adding a diffusion-type force contribution to the transport equation \cite{plasma}. The diffusion velocity, $\beta$, is introduced as a parameter. It controls the diffusion strength and typically corresponds to the velocity of electrons at the Fermi level. The permittivity tensor within this model can be written as:
\begin{equation}
    \frac{\overline{\epsilon}_{hydro}}{\epsilon_{0}}\left(\omega,\mathbf{k}\right)=\mathbf{1}+\frac{\omega_{p}^{2}}{\omega}\left[-\omega\mathbf{1}+i\omega_{c}\mathbf{\hat{z}}\times\mathbf{1}_{t}+\frac{\beta^{2}}{\omega}\mathbf{k}\otimes\mathbf{k}\right]^{-1}.
    \label{hydropermittivity}
\end{equation}
where we observe how the last term on the right hand side introduces an explicit dependence on the momentum.

Dispersion surfaces for this model are shown in Fig. 2(b), for a choice of parameters given in the figure caption with one pair of Weyl points, 
at $k_{z}=\pm k_{z}^W$, marked with a green dot in the figure. 
Cross sections of the dispersion, along the $k_{z}$ axis and at three transverse cuts are shown in Fig. 4. Only the longitudinal modes are affected by nonlocality, 
\cite{PhysRevB.92.125153}. This can be observed in the dispersion of modes along $k_z$ (panel a), where the flat longitudinal mode of the local model gains a positive group velocity ($\frac{\partial\omega}{\partial k}>0$, for positive $k$), and also in the in-plane dispersion (b-d, for different values of $k_{z}$), where the low-frequency TM mode now bends upward. This is an important detail because the high-frequency band gap is consequently eliminated in every cross section, for any $k_{z}$ value, in contrast to the results obtained with the local model (see Fig. \ref{local}(b)). 
The introduction of arbitrarily weak nonlocality of this type creates a type-II Weyl system, as can be attested by the hyperbolic isofrequency curves around the linear crossing seen in Fig. \ref{hydroWeyl}. This is the case when the crossing between the longitudinal mode and the transverse mode have group velocities $v_{g}=\frac{\partial\omega}{\partial k}$ with the same sign. This type of Weyl point has a diverging local density of states (DOS) at the Weyl frequency \cite{PhysRevLett.123.033901}.

Due to the closing of the high frequency band gap in planes transverse to the $k_z$ axis within this model, the Green's function method can only be applied to the low frequency band gap, highlighted in orange in Fig. 4. Direct computation of the gap Chern number for this band gap yields $\mathcal{C}_{gap}=0$, at any cross sectional plane, as seen for a plane in between the two Weyl points, $k_z=0$, in (b), a plane at the Weyl point, $k_z=k_z^W$ (c) and after the Weyl point, $k_z=\frac{3}{2}k_z^W$ (d). This is consistent with the fact that this band gap never closes, either with a Weyl crossing or any other type of degeneracy (see Fig. \ref{disp}(b)), such that the topological invariant of the gap cannot change. 

With the hydrodynamic model, we cannot then obtain information on the topological charge of the Weyl point, as it results from a crossing between high frequency bands (see Fig. 4) and the lack of full frequency band gap along transverse directions prevents the application of the Green's function method. We next consider a different and more general nonlocal model.    

\subsection{Full Cut-Off Model}

A general solution to regularize the topology of a continuum was introduced in Ref. \cite{PhysRevB.92.125153}, based on the introduction of a high-frequency spatial cut-off to the material response of a system, as:
\begin{equation}
     \overline{\epsilon}_{cut-off}\left(\omega,\mathbf{k}\right)=\epsilon_{0}\mathbf{1}+\frac{1}{1+k^{2}/k^{2}_{max}}\left(\overline{\epsilon}_{loc}(\omega)-\epsilon_{0}\mathbf{1}\right),
     \label{cutoffpermittivity}
\end{equation}
where $\overline{\epsilon}_{loc}$ stands for the original local response, $k^{2}=\mathbf{k}\cdot\mathbf{k}$ and $k_{max}$ models the high-frequency wave vector cut-off. For large values of the wave vector, $k\gg k_{max}$, the material response is suppressed, since for  $k\rightarrow\infty$, $\overline{\epsilon}_{cut-off}\left(\omega,\mathbf{k}\right) \rightarrow \epsilon_{0}\mathbf{1}$ and the material response becomes that of free space. 
\begin{figure}[ht]
    \centering
    \includegraphics[width=0.6\linewidth]{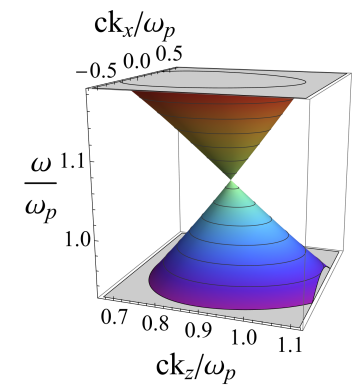}
    \caption{The full spatial cut-off model features Type I Weyl points: Dispersion surface and isofrequency curves around the outer Weyl point at $k_{z}=k_{z}^{W,2}$ in the $\hat{\mathbf{z}}$ and $\hat{\mathbf{x}}$ directions. The parameters used here were $k_{max}=\omega_{p}/c$, $\omega_{c}=0.8\omega_{p}$ and $\omega'_{p}=\sqrt{2}\omega_{p}$.}
    \label{cutoffWeyl}
\end{figure}
\begin{figure*}
    \centering
    \includegraphics[width=\textwidth]{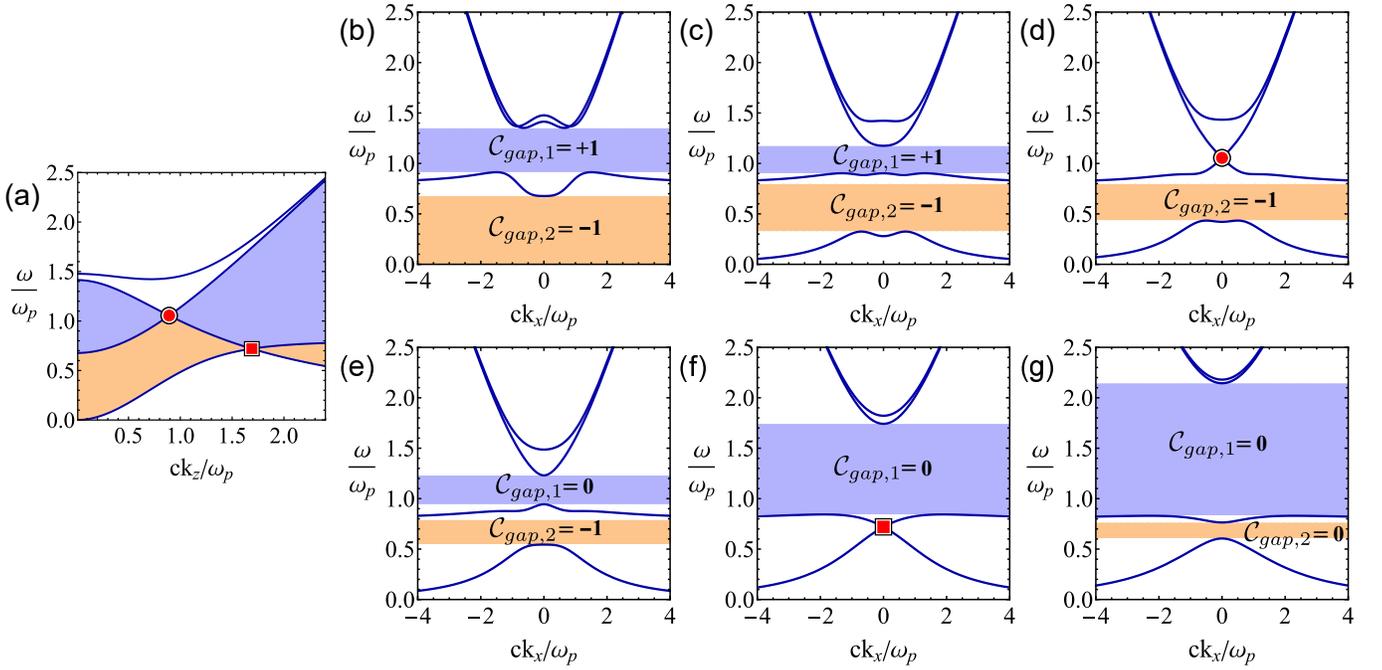}
    \caption{Full cut-off model: (a) Dispersion in the $k_{z}$ axis exhibiting the two regions of interest in blue and orange. (b)-(g) Cross sections of the 3D dispersion (FIG. \ref{disp}(c)) for different $k_{z}$ values: (b) $k_{z}=0$, (c) $k_{z}=\frac{3}{4}k_{z}^{W,1}$, (d) $k_{z}^{W,1}$, (e) $k_{z}=\frac{5}{4}k_{z}^{W,1}$, (f) $k_{z}^{W,2}$ and (g) $k_{z}=\frac{5}{4}k_{z}^{W,2}$. In (b)-(g), the high-frequency band gaps are highlighted in blue, the low-frequency ones in orange. The inner Weyl point at $k_{z}^{W,1}$ is highlighted with the red circle and the outer Weyl point with the red square at $k_{z}^{W,2}$.
    The parameters used here were $k_{max}=\omega_{p}/c$, $\omega_{c}=0.8\omega_{p}$ and $\omega'_{p}=\sqrt{2}\omega_{p}$.}
    \label{cutoff}
\end{figure*}

Unlike the hydrodynamic model, the nonlocality induced by this regularization affects all of the dispersion modes, albeit in a different way than the hydrodynamic model. Most importantly, the longitudinal mode along the $k_{z}$ axis again acquires a nonzero group velocity, but this time a negative one ($\frac{\partial\omega}{\partial k}<0$, for positive $k$), as can be observed in Fig. \ref{cutoff}(a). Additionally, the frequency of the longitudinal mode now tends to zero $\omega\to 0$ as $k\to\infty$ and consequently this mode always intersects the other two transverse modes. This originates two pairs of Weyl points whose momentum space locations will be designated as $k_{z}=\pm k_{z}^{W,1}$ for the inner crossing, and $k_{z}=\pm k_{z}^{W,2}$ for the outer crossing. The physics of these Weyl crossings is distinct from that of those appearing in hydrodynamic model, due to the different signs of the group velocities of each crossing mode. In this case, the isofrequency curves around all Weyl points are closed ellipses, with a vanishing DOS exactly at the Weyl frequency, making this a type-I Weyl system. This is shown in Fig. \ref{cutoffWeyl} for the outer Weyl point, $k_z=+k_{z}^{W,2}$. This confirms that nonlocality has a crucial role in determining the type and number of Weyl point exhibited in a system \cite{PhysRevLett.117.057401}.

We now consider the band structures in planes transverse to the $k_z$ axis, see Fig. \ref{cutoff}(b-g), in order to compute the topological charge of the Weyl points present in this system. As in the previous sections, we consider dispersion curves $\omega(k_x)$, for different values of $k_z$. As with the hydrodynamic model, we start at $k_z=0$ (b), where this time we identify two band gaps: a high frequency one, highlighted in blue, and a low frequency one, highlighted in orange. 
Computation of the gap Chern numbers yields $\mathcal{C}_{gap,1}=+1$ (blue band gap) and $\mathcal{C}_{gap,2}=-1$ (orange band gap). In contrast to the local magnetized plasma, the Chern numbers are now well defined due to the regularization provided by the spatial cut off, and in contrast to the hydrodynamic model, we observe two non-trivial band gaps. These results are consistent with those obtained in Ref. \cite{PhysRevB.92.125153}, where the magnetized plasma modes at $k_z=0$ were studied. 
Next, panel (c) shows the band dispersion away from $k_z=0$  but still at the left of the first Weyl point, $k_z=\frac{3}{4}k_z^{W,1}$. In this transverse plane, the zero-frequency mode is lifted, but the two band gaps are still preserved, as expected. Numerical computation confirms that the Chern numbers are still $\mathcal{C}_{gap,1}=+1$ and $\mathcal{C}_{gap,2}=-1$, as expected since the gaps have not closed. At the first Weyl point, $k_{z}=k_{z}^{W,1}$, the high frequency band gap closes at this linear degeneracy, as shown in panel (d), while the low frequency one is maintained, with $\mathcal{C}_{gap,2}=-1$, as expected too. Further increasing the value of $k_z$ to $k_z=\frac{5}{4}k_z^{W,1}$, in between the two Weyl points, we see in panel (e) how the high frequency gap opens up again, although the calculated gap Chern number is now $\mathcal{C}_{gap,1}=0$. At the same time, the lower frequency band gap does not close in this process and hence $\mathcal{C}_{gap,2}=-1$ is kept. This band gap only closes in panel (f), at the second Weyl point, $k_z=k_z^{W,2}$. Finally, for $k_z$ at the right of this outer degeneracy, we find this band gap is open again (g), and its gap Chern number turns trivial, $\mathcal{C}_{gap,2}=0$, such that all band gaps are now trivial. This is expected as this should be the behaviour at $k\rightarrow\infty$, and there are no more band gap closings after the outer Weyl point that could enable a change in topological invariants. 

Once we know all the gap Chern numbers, we can calculate the 
magnitude of the topological charge of the Weyl point using Eq. (\ref{charge}). We obtain, for the inner Weyl point at 
positive momentum, $\mathcal{C}_{W}(k_z=+k_z^{W,1})=0-(+1)=-1$, while its counterpart at negative momentum has charge $\mathcal{C}_{W}(k_z=-k_z^{W,1})=+1-0=+1$. This value pf the topological charge is consistent with the one obtained in the local model, as should be expected.
Additionally, the topological charges of the outer Weyl pair are $\mathcal{C}_{W}(k_z=+k_z^{W,2})=0-(-1)=+1$ for the Weyl crossing with positive momentum and $\mathcal{C}_{W}(k_z=-k_z^{W,2})=(-1)-0=-1$ for the crossing with negative momentum. The outer pair of Weyl points smoothly moves to infinity by increasing $k_max$, thus retrieving the results from the local model. 


\begin{figure}
    \centering
    \includegraphics[width=0.4\textwidth]{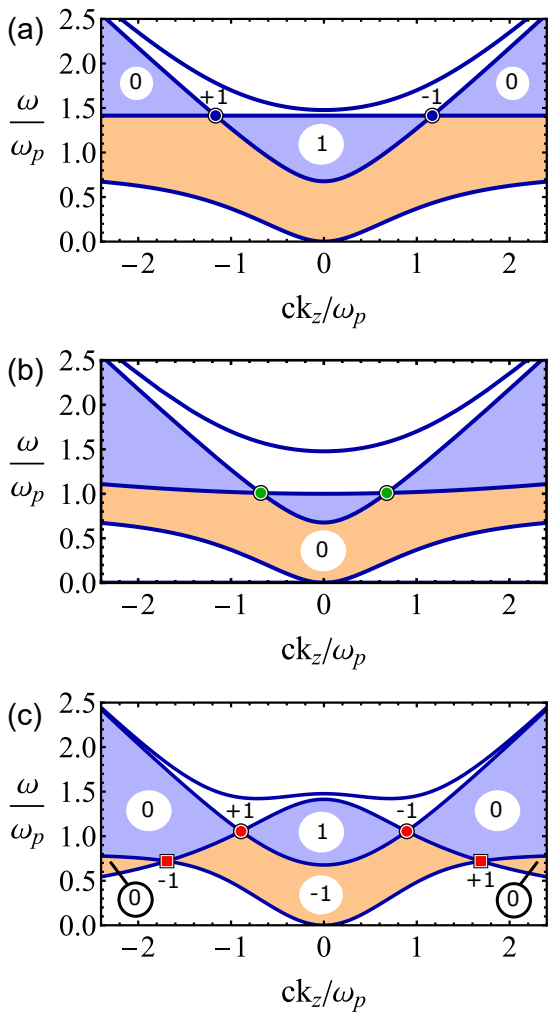}
    \caption{Summary of the topological characterizations of the local and nonlocal models of the 3D magnetized plasma. The computed gap Chern numbers in transverse cuts are highlighted with a white circle, for each subregion. The topological charges of the Weyl points are also showcased. In the three plots we see the $k_{z}$ axis dispersion for: (a) local model with the Weyl pair as blue circles. Parameters used were $\omega_{c}=0.8\omega_{p}$ and$\omega'_{p}=\sqrt{2}\omega_{p}$; (b) hydrodynamic model with the Weyl pair as green circles. The parameters used were $\beta=0.2c$, $\omega_{c}=0.8\omega_{p}$ and $\omega'_{p}=\omega_{p}$; (c) full cut-off model with the inner Weyl pair as red circles and the outer pair as red squares.  The parameters used were $k_{max}=\omega_{p}/c$, $\omega_{c}=0.8\omega_{p}$ and $\omega'_{p}=\sqrt{2}\omega_{p}$.}
    \label{results}
\end{figure}

\section{Conclusions}
In this work we have introduced a first principles formalism that enables the characterization of Weyl points in 3D dispersive photonic continua. In particular, the topological charge of Weyl points is obtained through the calculation of gap Chern numbers in planes transverse to the axis that joins a Weyl pair, by means of the photonic Green's function of the medium \cite{PhysRevB.92.125153}. This is in contrast to standard methods based on the direct computation of the Berry curvature, which require the calculation of the eigenvectors at each value of the wave vector and are thus computationally intensive. 

We have shown an application of the theory for a 3D magnetized plasma, where the $\mathcal{T}$-breaking enables the existence of Weyl point pairs. 
When considering a standard local description of the magnetized plasma, a Weyl point pair emerges at the crossing between a flat plasmonic longitudinal mode and a transverse mode. However, it is known that dispersive photonic media possess ill-defined topology, so we have also considered two response models that regularize the topology. First, the hydrodynamic model introduces a non-zero diffusion velocity for the electrons that accounts for electron-electron repulsion. This results in a bending of the longitudinal mode by bending it upward, which changes the physical properties of the Weyl points. In particular, the isofrequency curves around the Weyl points of this model are hyperbolic, such that they constitute Type-II Weyl points. Second, we have considered a general regularization method based on introducing a full spatial cut-off. In this case the longitudinal mode bends downwards, and two Weyl point pairs emerge. Additionally, they present closed elliptical isofrequency curves, such that they are Type-I. This shows that introducing nonlocality is needed to regularize the topology of photonic continua, but also that it is important to choose carefully the particular model to implement nonlocality. Furthermore, the number and type of Weyl points depends on the nonlocality model, even if the parameter that controls the nonlocal effects is arbitrarily small.

The topological charges of Weyl points were calculated by applying the Green's function method in cross-sectional planes, from the difference between the gap Chern number of the relevant frequency band gaps at each side of the Weyl points. Results for the three considered models are summarized in Fig. \ref{results}, with the Weyl charge marked at the side of each Weyl point, and the gap Chern numbers given within the gaps. First, for the local model (results show in a) only the topology of the high frequency band gap is well defined, and a gap Chern number can only be attributed to this band gap (shaded in blue) and not to the low frequency one. Calculations yield a gap Chern number of 0 before the Weyl point at negative momentum, of 1 in between the Weyl points, and 0 after the Weyl point at positive momentum. From this topological charges of $+1$ and $-1$ are obtained for the two Weyl points, consistent with the fact that one should be a source and the other a drain of Berry flux. Second, a magnetized plasma described by a hydrodynamic model also shows a Weyl pair (see panel b), although the lack of full band gaps in the transverse direction prevents us from applying our method and calculating gap Chern numbers. Finally, a full spatial cut-off approach to regularize the model enables the existence of two Weyl pairs surrounded by full band gaps in the transverse plane, such that we can fully characterize this model. As shown in panel (c), starting at negative values of $k_z$, the two relevant band gaps have trivial gap Chern numbers. Each turns non-trivial, with gap Chern numbers of $-1$ for the low frequency band gap, and $+1$ for the high frequency band gap), after the outer and inner Weyl points are crossed, respectively. Crossing the counterparts of the Weyl pairs changes both gap Chern numbers back to 0, consistently with the trivial response that should be obtained at high wave vectors. This allows us to conclude that the outer Weyl pair has topological charges of $-1$ and $+1$ for negative and positive momentum, respectively, while the inner pair has charges of $+1$ and $-1$.  

In conclusion, we have presented a detailed study of 3D gapless phases in dispersive photonic continua, and discussed the important effects of nonlocality in the topological properties of these systems. Specifically, we made use of the link between the topological charge of the 3D Berry curvature singularities (Weyl points) and the gap Chern numbers of 2D sub-systems, to characterize the topological properties of an electromagnetic continuum in 3D for the first time. Our results are based on first principle calculations of topological invariants that naturally take into account frequency dispersion and nonlocality. This methodology can be extended to $\mathcal{P}-$broken Weyl systems, other 3D topological phases, or to Floquet systems. 

\begin{acknowledgments}
We acknowledge funding from Funda\c c\~ao para a Ci\^encia e a Tecnologia and Instituto de Telecomunica\c c\~oes under projects UIDB/50008/2020, 2022.06797.PTDC and UTAP-EXPL/NPN/0022/2021. PAH acknowledges financial support from the Spanish Ministry of Science and Innovation, through the Ram\'on y Cajal program (Grant No. RYC2021-031568-I) and from the María de Maeztu Programme for Units of Excellence in R\&D (CEX2018-000805-M), and from the CAM (Y2020/TCS-6545). MGS acknowledges support from the Institution of Engineering and Technology (IET) and by the Simons Foundation (Award: 733700).
\end{acknowledgments}

\appendix


\section{Frequency-Independent Differential Operator $\hat{L}_{\mathbf{k}}$}

Here we show the derivation of the operator $\hat{L}$ from which $\hat{L}_{\mathbf{k}}$ is obtained. We introduce a general method which is described in \cite{PhysRevLett.129.133903} and we consider the lossless hydrodynamic model of a magnetized plasma, as a first instance.

We start with Maxwell's equations in time domain, for propagation in free space ($\partial_{t}\equiv\frac{\partial}{\partial_{t}}$):
\begin{equation}
\begin{split}
     -i\nabla\times\mathbf{E}&=i\mu_{0}\partial_{t}\mathbf{H}, \\
     i(\nabla\times\mathbf{H}-\mathbf{j})&=i\epsilon_{0}\partial_{t}\mathbf{E},
    \end{split}
    \label{iMaxwell}
\end{equation}
where $\mu_{0}$ and $\epsilon_{0}$ are the vacuum's permeability and permittivity, respectively.
We then consider the continuity equation given by:
\begin{equation}
    \partial_{t}\rho+\nabla\cdot\mathbf{j}=0.
    \label{continuity}
\end{equation}

The current $\mathbf{j}$ and charge $\rho$ densities model the response of the dispersive electric gyrotropic material. Making use of Newton's second law of motion plus Lorentz's Force law, the transport equation for a free electron gas biased with a static magnetic field $(\mathbf{B}_{0}=B_{0}\hat{z})$ is given by:
\begin{equation}
    \frac{d\mathbf{j}}{dt}=\epsilon_{0}\omega_{p}^{2}\mathbf{E}+\frac{q}{m}\mathbf{j}\times\mathbf{B}_{0}-\beta^{2}\nabla\rho,
    \label{hydrotransport}
\end{equation}
where $q=-e$ is the charge of an electron with opposite sign and $m$ is its effective mass. If we wish to retrieve the local model, we just set $\beta$ as zero.
Equations (\ref{iMaxwell})-(\ref{hydrotransport}) can be rewritten as a Schr\"{o}dinger-type equation:
\begin{equation}
    \hat{L}\cdot\mathbf{Q}=\frac{1}{c}i\partial_{t}\mathbf{Q}.
    \label{eigenvalueequ}
\end{equation}

The state vector $\mathbf{Q}$ is given by $\mathbf{Q}=\left[E_{x}\quad E_{y}\quad E_{z}\quad \Tilde{H}_{x}\quad \Tilde{H}_{y}\quad \Tilde{H}_{z}\quad \Tilde{j}_{x}\quad \Tilde{j}_{y}\quad \Tilde{j}_{z}\quad \Tilde{\rho}\right]^{T}$. We introduce the normalized magnetic field, current and charge density: $\Tilde{\mathbf{H}}=\eta_{0}\mathbf{H}$, $\Tilde{\mathbf{j}}=\eta_{0}\mathbf{j}$, $\Tilde{\rho}=\eta_{0}c\rho$, with $\eta_{0}$ being the vacuum wave impedance and c the speed of light. Since we are interested in the 3D case of the continuous magnetized plasma, we assume a spatial and time variation of the state vector of the type $e^{i\mathbf{k}\cdot\mathbf{r}}e^{-i\omega t}$, with wave vector $\mathbf{k}=k_{x}\hat{x}+k_{y}\hat{y}+k_{z}\hat{z}$. We are also interested in analysing the dispersion of all light modes in this medium, and hence the fields and wave vector are unconstrained: $\mathbf{E}=E_{x}\hat{x}+E_{y}\hat{y}+E_{z}\hat{z}$, $\mathbf{H}=H_{x}\hat{x}+H_{y}\hat{y}+H_{z}\hat{z}$ and $\mathbf{j}=j_{x}\hat{x}+j_{y}\hat{y}+j_{z}\hat{z}$ which justifies the dimension of the state vector written above. By manipulating equations (\ref{iMaxwell})-(\ref{hydrotransport}), and taking into account that the cyclotron frequency is defined as $\omega_{c}=-qB_{0}/m$, we derive the operator $\hat{L}$:
\begin{widetext}
\begin{equation}
    \hat{L}(-i\nabla)=
    \begin{pmatrix}
    0 & 0 & 0 & 0 & -i\partial_{z} & i\partial_{y} & -i & 0 & 0 & 0 \\
    0 & 0 & 0 & i\partial_{z} & 0 & -i\partial_{x} & 0 & -i & 0 & 0 \\
    0 & 0 & 0 & -i\partial_{y} & i\partial_{x} & 0 & 0 & 0 & -i & 0 \\
    0 & i\partial_{z} & -i\partial_{y} & 0 & 0 & 0 & 0 & 0 & 0 & 0 \\
    -i\partial_{z} & 0 & i\partial_{x} & 0 & 0 & 0 & 0 & 0 & 0 & 0 \\
    i\partial_{y} & -i\partial_{x} & 0 & 0 & 0 & 0 & 0 & 0 & 0 & 0 \\
    i\frac{\omega_{p}^{2}}{c^{2}} & 0 & 0 & 0 & 0 & 0 & 0 & -i\frac{\omega_{c}}{c} & 0 & -i\frac{\beta^{2}}{c^{2}}\partial_{x} \\
    0 & i\frac{\omega_{p}^{2}}{c^{2}} & 0 & 0 & 0 & 0 & i\frac{\omega_{c}}{c} & 0 & 0 & -i\frac{\beta^{2}}{c^{2}}\partial_{y} \\
    0 & 0 & i\frac{\omega_{p}^{2}}{c^{2}} & 0 & 0 & 0 & 0 & 0 & 0 & -i\frac{\beta^{2}}{c^{2}}\partial_{z} \\
    0 & 0 & 0 & 0 & 0 & 0 & -i\partial_{x} & -i\partial_{y} & -i\partial_{z} & 0 
    \end{pmatrix}.
    \label{Lhydro}
\end{equation}

The operator $\hat{L}_{\mathbf{k}}$ from equation (\ref{Green}) can be obtained by substituting the spatial derivatives in (\ref{Lhydro}) with the corresponding wave vector components $\frac{\partial}{\partial j}\leftrightarrow ik_{j}$, $j=\{x,y,z\}$. To get the local model analysed in this work we must set $\beta$ to zero and include the uniaxial plasma frequency component $\omega'_{p}$.

The full cut-off model can be enforced in the system by changing the current density vector $\mathbf{j}$ in Maxwell's equations (\ref{iMaxwell}), replacing it with $(-k^{-2}_{max}\nabla^{2}+1)^{-1}\mathbf{j}$. The derived operator for this model with the uniaxial plasma frequency $\omega'_{p}$ is given by:

\begin{equation}
\begin{split}
    \hat{L}(-i\nabla)= \begin{pmatrix}
    0 & 0 & 0 & 0 & -i\partial_{z} & i\partial_{y} & \frac{-i}{(-k^{-2}_{max}\nabla^{2}+1)} & 0 & 0 & 0 \\
    0 & 0 & 0 & i\partial_{z} & 0 & -i\partial_{x} & 0 & \frac{-i}{(-k^{-2}_{max}\nabla^{2}+1)} & 0 & 0 \\
    0 & 0 & 0 & -i\partial_{y} & i\partial_{x} & 0 & 0 & 0 & \frac{-i}{(-k^{-2}_{max}\nabla^{2}+1)} & 0 \\
    0 & i\partial_{z} & -i\partial_{y} & 0 & 0 & 0 & 0 & 0 & 0 & 0 \\
    -i\partial_{z} & 0 & i\partial_{x} & 0 & 0 & 0 & 0 & 0 & 0 & 0 \\
    i\partial_{y} & -i\partial_{x} & 0 & 0 & 0 & 0 & 0 & 0 & 0 & 0 \\
    i\frac{\omega_{p}^{2}}{c^{2}} & 0 & 0 & 0 & 0 & 0 & 0 & -i\frac{\omega_{c}}{c} & 0 & 0 \\
    0 & i\frac{\omega_{p}^{2}}{c^{2}} & 0 & 0 & 0 & 0 & i\frac{\omega_{c}}{c} & 0 & 0 & 0 \\
    0 & 0 & i\frac{\omega_{p}'^{2}}{c^{2}} & 0 & 0 & 0 & 0 & 0 & 0 & 0 \\
    0 & 0 & 0 & 0 & 0 & 0 & -i\partial_{x} & -i\partial_{y} & -i\partial_{z} & 0 
    \end{pmatrix}.
    \label{Lcut-off}
\end{split}
\end{equation}
\end{widetext}

\section{Uniaxial Plasma Response}

\begin{figure}[H]
    \centering
    \includegraphics[width=0.50\textwidth]{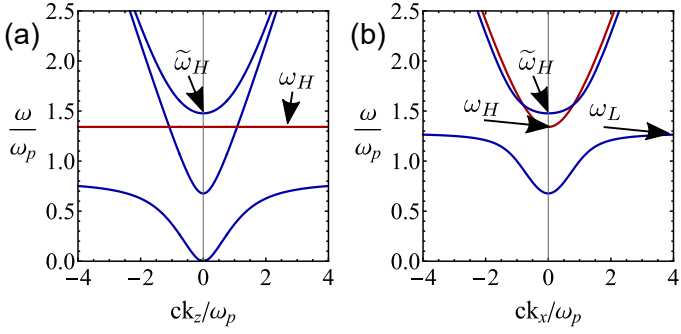}
    \caption{Frequency dispersion for the local model of a magnetized plasma, along (a) the $k_{z}$ axis with the longitudinal mode highlighted in red and along (b) the $k_{z}$ axis with the TE mode highlighted in red. The longitudinal mode frequency (a), high-frequency band minimum (a), TE band minimum (b) and low-frequency TM band high wave vector limit (b) are all pointed out in the figure. The parameters used here were $\omega_{c}=0.8\omega_{p}$ and $\omega'_{p}=\sqrt{2}\omega_{p}$.}
    \label{uniaxial}
\end{figure}

In this section, we discuss the uniaxial plasma response in the $\mathbf{\hat{z}}$ direction. As stated in the main text, this response allows for full band gaps away from both Weyl points, that ultimately allows for the calculation of their topological charges with our method. Without it, the high-frequency band gaps (shaded in blue in Fig. \ref{local} and Fig. \ref{cutoff}) present in the local and full cut-off models would be non-existent. Here we provide a range of values for $\omega'_{p}$ that guarantees a full band gap before and after the relevant Weyl point.
Firstly, we may define the uniaxial plasma frequency $\omega'_{p}$ as a function of $\omega_{p}$ as $\omega'_{p}=\sqrt{K}\omega_{p}$, where $K$ is a scalar. Consider the $xoy$ plane dispersion of the local model in figure \ref{uniaxial}(b) featuring both TM modes in blue and one TE mode in red. The dispersion relations for the photonic modes are:
\begin{equation}
    k^{2}=\epsilon_{z}\left(\frac{\omega}{c}\right)^{2},\quad \text{TE modes},
    \label{TEdisplasma}
\end{equation}
\begin{equation}
    k^{2}=\frac{\epsilon_{t}^{2}-\epsilon^{2}_{g}}{\epsilon_{t}}\left(\frac{\omega}{c}\right)^{2},\quad \text{TM modes}.
    \label{TMdisplasma}
\end{equation}
Using the explicit expressions of the constitutive parameters, the following solutions are derived:
\begin{equation}
    \omega_{TE}=\pm\sqrt{c^{2}k^{2}+\omega'^{2}_{p}},
    \label{TEdissol}
\end{equation}
\begin{eqnarray}
    &&\omega_{TM}=\\ \nonumber &&\pm\sqrt{\frac{c^{2}k^{2}+\omega^{2}_{c}+2\omega^{2}_{p}\pm\sqrt{c^{4}k^{4}-2c^{2}k^{2}\omega^{2}_{c}+\omega^{4}_{c}+4\omega^{2}_{c}\omega^{2}_{p}}}{2}}.
    \label{TMdissol}
\end{eqnarray}
Let us define $\Delta\omega=\omega_{H}-\omega_{L}$ as the function that measures the amplitude of the high-frequency band gap between the low-frequency TM mode and the TE mode, with $\omega_{H}=\omega'_{p}/\omega_{p}$ and $\omega_{L}=\left(\sqrt{\omega_{c}^{2}+\omega^{2}_{p}}\right)/\omega_{p}$. These values correspond to the lowest frequency limit of the TE mode and the highest frequency limit of the low-frequency TM mode, respectively. In order to obtain a full band gap, we need to find $K$ that satisfies $\Delta\omega>0$. Solving the inequality gives us a lower bound for this parameter:
\begin{equation}
    K>\frac{\omega^{2}_{c}}{\omega^{2}_{p}}+1.
    \label{TETMgapC}
\end{equation}

Consider now the $k_{z}$ axis dispersion of the local model in Fig. \ref{uniaxial}(a), featuring the flat longitudinal mode with constant frequency $\omega=\omega'_{p}$ in red whose intersection with the transverse mode in blue originates the Weyl pair. Naturally, the longitudinal mode is also affected by this response, shifting it upwards. If the plasma frequency $\omega'_{p}$ is too large, the longitudinal mode will intersect the high-frequency mode (in blue as well), potentially creating new Weyl points, thus altering the topology of the system, which is not our objective. We must then set an upper bound in the form $\omega'_{p}<\widetilde{\omega}_{H}$, where $\widetilde{\omega}_{H}$ is the lowest frequency limit of the high-frequency transverse mode in $k_{z}$ but also the lowest frequency limit of the high-frequency TM mode in the xoy plane. The upper bound for this parameter must then be:
\begin{equation}
     K<\frac{\omega_{c}^{2}}{2\omega_{p}^{2}}+\frac{\omega_{c}}{\omega_{p}^{2}}\sqrt{\left(\frac{\omega_{c}}{2}\right)^2+\omega^{2}_{p}}+1.
\end{equation}

The following equation represents the bounds of the interval for which the $K$ parameter does not cause the emergence of new band degeneracies but also allows the existence of full high-frequency band gaps in the cross sections, depending only on $\omega_{c}$ and $\omega_{p}$:
\begin{equation}
    \frac{\omega^{2}_{c}}{\omega^{2}_{p}}+1<K<\frac{\omega_{c}^{2}}{2\omega_{p}^{2}}+\frac{\omega_{c}}{\omega_{p}^{2}}\sqrt{\left(\frac{\omega_{c}}{2}\right)^2+\omega^{2}_{p}} +1,
    \label{condition}
\end{equation}
which justifies our choice of $\omega'_{p}=\sqrt{2}\omega_{p}$ in the local and full cut-off models, for $\omega_{c}=0.8\omega_{p}$.

\section{Negative Frequency Modes}

\begin{figure}[H]
    \centering
    \includegraphics[width=0.50\textwidth]{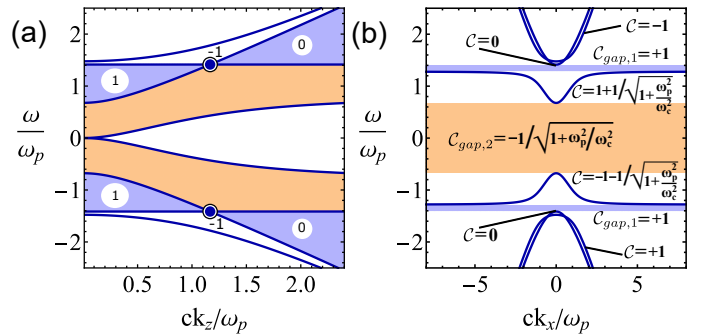}
    \caption{Frequency dispersion for a local magnetized plasma, featuring negative frequency modes. (a) $k_{z}$ axis exhibiting the positive momentum Weyl points marked with a blue circle and their topological charge. The band gaps of interest are shaded in blue and orange. The computed gap Chern numbers in transverse cuts are highlighted with a white circle, for each subregion. (b) $xoy$ plane ($k_z=0$), with Chern numbers for each gap ($\mathcal{C}_\text{gap}$) and for each band ($\mathcal{C}$) in the insets. The parameters used here were $\omega_{c}=0.8\omega_{p}$ and $\omega'_{p}=\sqrt{2}\omega_{p}$.}
    \label{neglocal}
\end{figure}

The importance of the bands with negative frequency to the band gaps' topology is highlighted in this section. We start by illustrating why the ill-defined topology \cite{PhysRevB.97.115146,Chern,sym13112229,PhysRevLett.129.133903} of the TM mode in the $xoy$ plane of the local magnetized plasma does not affect the topology of the high-frequency band gap, even though the gap Chern number is a sum of the individual Chern numbers of the bands below the band gap. Figure \ref{neglocal}(b), shows the gap Chern numbers $\mathcal{C}_\text{gap}$ and the band Chern numbers $\mathcal{C}$ in the $xoy$ plane ($k_{z}=0$) of the local model. The gap Chern numbers are frequency-symmetric $\mathcal{C}_\text{gap}(\omega)=\mathcal{C}_\text{gap}(-\omega)$, while the band Chern numbers of symmetric modes have opposite signs. The latter property explains why the high-frequency band gap (shaded in blue) in the $xoy$ plane of the local model has a well-defined topology, because the non-integer Chern numbers of the low-frequency TM mode and its negative counterpart cancel each other. 
\begin{figure}[H]
    \centering
    \includegraphics[width=0.5\textwidth]{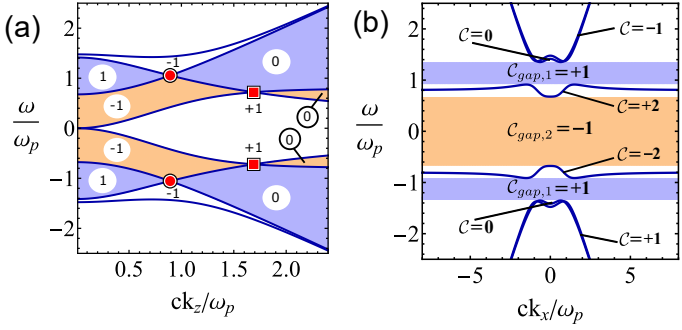}
    \caption{Frequency dispersion for a full cut-off model of a  magnetized plasma, featuring negative frequency modes. (a) $k_{z}$ axis exhibiting the positive momentum Weyl points marked with a red circle for the inner ones, with a red square for the outer ones and their topological charge. The band gaps of interest are shaded in blue and orange. The computed gap Chern numbers in transverse cuts are highlighted with a white circle, for each subregion. (b) $xoy$ plane ($k_z=0$), with Chern numbers for each gap ($\mathcal{C}_\text{gap}$) and for each band ($\mathcal{C}$) in the insets. The parameters used here were $k_{max}=\omega_{p}/c$, $\omega_{c}=0.8\omega_{p}$ and $\omega'_{p}=\sqrt{2}\omega_{p}$.}
    \label{negcutoff}
\end{figure}
A similar analysis of the full frequency spectrum is done for the full cut-off model in Fig. \ref{negcutoff}. Since the gap Chern numbers are even-symmetric with respect to frequency, Weyl points with equal momentum but opposite frequency values have identical topological charge. This property is shown in panel (a) of both pictures. The band Chern numbers are the analytical results from \cite{PhysRevB.97.115146} and they were computed with the theory from \cite{PhysRevB.92.125153}.

\bibliographystyle{apsrev4-1}
\bibliography{apssamp}

\end{document}